\documentclass[conference]{IEEEtran}
\IEEEoverridecommandlockouts
\usepackage{cite}
\usepackage{amsmath,amssymb,amsfonts}
\usepackage{algorithmic}
\usepackage{graphicx}
\usepackage{textcomp}
\usepackage{xcolor}
\def\BibTeX{{\rm B\kern-.05em{\sc i\kern-.025em b}\kern-.08em
    T\kern-.1667em\lower.7ex\hbox{E}\kern-.125emX}}
\begin{document}

\title{Distributed Autonomous Organizations as Public Services Supplying Platform \\
}

\author{\IEEEauthorblockN{1\textsuperscript{st} Giovanni De Gasperis}
\IEEEauthorblockA{\textit{DISIM} \\
\textit{Università degli Studi dell'Aquila}\\
L'Aquila, Italy \\
0000-0001-9521-4711}
\and
\IEEEauthorblockN{2\textsuperscript{nd} Sante Dino Facchini}
\IEEEauthorblockA{\textit{DISIM} \\
\textit{Università degli Studi dell'Aquila}\\
L'Aquila, Italy \\
0000-0002-2009-5209}
\and
\IEEEauthorblockN{3\textsuperscript{rd} Maurizio Michilli}
\IEEEauthorblockA{\textit{CEO} \\
\textit{Servizi Elaborazione Dati Spa}\\
L'Aquila, Italy \\
m.michilli@sedaq.it}


}

\maketitle

\begin{abstract}
Servizi Elaborazioni Dati SpA is a public company owned by Municipality of L'Aquila, it supplies the institution with network services and software applications for distributing services to citizens. The future policy of the company is to enlarge the offer of its services to nearby communities that are unable to set up and maintain their own network and software structures. This paper presents thus a possible architecture model to support small municipalities in supplying public services to citizens, with the aid of SED Spa. Through second level platforms based on Blockchain networks and Multi-agents Systems running on smart contracts, the system will focus on Waste Tax (Ta.Ri) management system in the Fascicolo del Cittadino environment.
\end{abstract}


\section{Introduction}
Italian government has recently introduced the Digital Administration Code (CAD) a law that brings together all the best practices concerning the modernization of the Public Administration. It focuses on interactions with citizens and businesses in order to promote and make the rights of digital citizenship effective, as per  DL No.217/2017\footnote{https://www.gazzettaufficiale.it/eli/id/2018/1/12/18G00003/sg}.

A first pillar is the equalization of the legal validity of  digital documents to paper ones, this also implies the acknowledgment of the electronic or digital signature as fully equivalent the traditional one.
Even more important is that digitalization makes possible to create a circuit of public administrations to easily communicate with each other on heterogeneous networks, speaking a unambiguous and common language recognised by all players. 
All the innovation processes are supervised by a technical agency Agenzia per l'Italia Digitale (AgID) that introduced tools like Digital Signature, a double key asymmetric encription system for watermarking digital documents, SPID -Sistema Pubblico per l'Identità Digitale- a certified account released by technical authorized partners of AgID and PagoPA a payment system to fast and securely pay taxes, fines and fees to Public Administration entities. All this technologies converges into the Fascicolo del Cittadino, a digital portfolio of documents, paperworks and bills that connect the citizen directly to his local municipality. 
In this scenario local administrations play an essential role being the first point of contact with citizens for many e-government services. 

\section{A distribution model for small communities}
\subsection{Actual SED Architecture}
SED - Servizi Elaborazione Dati SpA\footnote{http://www.sedaq.it} is an in-house providing public shareholder company, fully controlled by the Municipality of L'Aquila\footnote{https://www.comune.laquila.it} and established in January 2001 as a spin-off from Azienda Servizi Municipalizzati.SED SpA falls within the category of public companies affected by Law No.135/2012.
Main services regards production, management and maintenance of municipal tax and general registry databases and applications, support for tax and extra-tax revenues billing, support activities in tax management and revenue assessment and collection. SED also collaborates with the Special Office for Restoration designing and managing applications and databases related to the 2009 Abruzzo earthquake.

\subsection{Problems of Small Municipalities in Disadvantaged Areas}
L'Aquila is the capital city of Abruzzo Region and the biggest municipality of Internal areas. The district is a mountainous territory distinguished for being scarcely populated but very vast, the population is in fact distributed in 108 municipalities with the majority living in small cities and villages under 5,000 inhabitants.
This spread configuration of very small communities brings many problems in term of guaranteeing a uniform service, most of municipalities in fact doesn't have a sufficient tax income for developing and setting up their own application for services such as waste collection or house tax billing. Many of them have teamed up for services such as local police or school buses, but when it comes to ICT and networking services higher set up and maintenance costs are a real unaffordable burden.
In this way municipalities must rely on external private companies to implement services like Property Tax collection or Register Office. Our case study will focus on Ta.Ri. waste tax management as an informal survey taken by SED pointed out it is a service that most of the municipalities still do in house with small automation.

\subsection{A Smart City Model Extended to Nearby Small Municipalities}
The idea of SED Spa is very simple: export models and applications already in place for Municipality of L'Aquila to smaller nearby administrations. This would generate on one side an increment of the company turnover, but would also represent a chance for smaller cities to benefit of the economy of scale by sharing services and avoiding the initial costs of developing their own applications and infrastructures.
This extension of services requires to carefully address problems like trusting partners, traceability of transactions and messaging, immutability of data and identity verification. The aim of this abstract is to propose an architecture to meet all this features and functionalities through the applications of Blockchain and DAO technologies as described in the next paragraphs\cite{SmartCities}. 

\section{Technologies and Architecture}
\subsection{Distributed Autonomous Organizations}
The definition of Distributed Autonomous Organizations (DAOs) has been formalized during the 90s and was associated to the use of Multi-agent Systems for intelligent home sensors control \cite{Dilger}. After 2008, with the introduction of the Digital Autonomous Corporations idea (DACs), a representation of a real company on the blockchain was defined; taking advantage from the introduction of Ethereum instruments like tokenization of shares, automated and unalterable smart contracts and transparent transactions register were finally applicable to the real governance process of an institution\cite{Buterin}.
DAOs as we know them today are an evolution of DACs, technically them refer to a system for modeling an organization through the deployment of interacting smart contracts based on a underlying blockchain network. They also get the properties of the blockchain layers such as absence of centralized control, security with cryptographic keys and self-executable smart contracts. All these functionalities make DAOs a credible environment to operate a virtual organization modeling all functions and procedures of a real one\cite{Chohan}.

\subsection{An Architecture for Distributing Public Services}\label{AA}
SED, Municipality of L'Aquila and small nearby Municipalities can be seen as components of the same Distributed Organization and thus as nodes of a blockchain with a preset group of Smart Contracts, Fungible and/or Non Fungible Tokens.  A possible framework to develop our architecture is Aragon\footnote{https://aragon.org/}, a second level platform based on  Ethereum\footnote{https://ethereum.org/} blockchain network designed to simulate government systems of public and private entities. It is easy to use and allows to deploy trial application at very cheap costs using Ethereum’s test nets such as Rinkeby\footnote{https://www.rinkeby.io/}. 

\begin{figure}[htbp]
\centerline{\includegraphics[width=0.75\columnwidth]{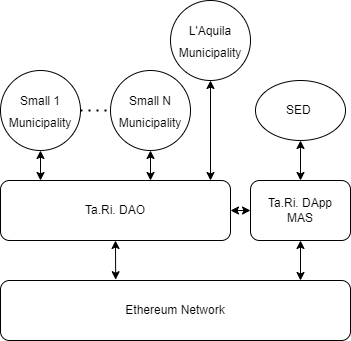}}
\caption{Ta.Ri dApp and DAO.}
\label{fig}
\end{figure}
Aragon stack is composed of: an Operative system environment where applications are abstracted from the underlying layer, a Packet Manager to distribute different versions of the software, APIs to manage requests on transactions, status and software without depending on a centralised service and a toolkit for the development of the dApps user interface. 
The Ta.Ri. app would thus be designed and deployed as a distributed app on Ethereum and implemented as a set of smart contracts running on Municipalities nodes. It would then interact with Municipalities users through the DAO agent module, a special app to interface external smart contracts. In this way the Aragon Ta.Ri-DAO will offer, among other standard applications like voting and maintainance ones, an internal app for Ta.Ri. billing, control and management. The dApp may include a Multi-agent system in order to perform behavioural analysis of actors involved in the DAO as well as prediction of tax income connected to other financial and demographic parameters\cite{Degas}.

\subsection{Measuring Results}
The project has as main key performance indicators the number of small municipalities that adopt the application and the number of citizens involved. The goal is to involve at least L'Aquila and two other Municipalities for a minimum of 80,000 citizens.



\end{document}